\documentclass[10pt,journal,compsoc]{IEEEtran}

\usepackage{comment}
\usepackage{color}
\usepackage{graphicx}
\usepackage{amsmath}
\usepackage{epstopdf}
\usepackage{multicol}
\usepackage{multirow}
\usepackage{multicol}

\begin{document}

\title{Hunting for supernovae articles\\in the universe of scientometrics\thanks{The Google Scholar's data presented in the article were collected on December 10th, 2019.}}

\author{
   Dimitrios Katsaros
}

\IEEEtitleabstractindextext{
\begin{abstract}
This short note records an unusual situation with some Google Scholar's profiles that imply the existence of ``supernovae'' articles, i.e., articles whose
impact -- in terms of number of citations -- in a single year gets (almost) an order of magnitude higher than the previous year and immediate drops (and remains
steady) to a very low level after the next year. We analyse the issue and resolve the situation providing an answer whether there exist supernovae articles.
\end{abstract}
\begin{IEEEkeywords}
supernovae articles, citations, Google Scholar, scientometrics.
\end{IEEEkeywords}
}

\maketitle

\IEEEpeerreviewmaketitle

\newcommand{\minitab}[2][l]{\begin{tabular}{#1}#2\end{tabular}}

\section{Introduction}
\label{sec-intro}

``A supernova is the biggest explosion that humans have ever seen. Each blast is the extremely bright, super-powerful explosion of a star. These spectacular 
events can be so bright that they outshine their entire galaxies for a few days or even months."\footnote{Definition and descriptions taken from https://spaceplace.nasa.gov/supernova/en/.}
The {\it progenitor} (i.e., the original object) -- depending on its size -- will eventually turn into a neutron star (or black hole) or it will vanish
completely by dispersing its remnants into the space. In any case, the qualitative outcome is that its after-explosion luminosity will be orders of magnitude
lower than that of the explosion, and this luminosity degradation will happen within a few months!

So, drawing on the analogy of supernovae in astronomy, we ask whether {\it there are supernovae scientific article(s)}, i.e., articles which have received a 
significant amount of citations for very short periods of time, say a few months, and then their impact -- in terms of citations -- vanishes.

\section{Do supernovae articles exist?}
\label{sec-negation}

During the last two decades the availability of rich bibliometric data in online databases such as Google Scholar, Elsevier Scopus, Clarivate Analytics' WoS 
allows for the data-centric study of the performance of various entities participating and shaping the research landscape. These bibliographic databases are used 
extensively to record the performance of scientists, and it is very common that scientists applying for promotion or funding refer to their profile in these 
services in order to prove their productivity and/or impact. Therefore, both the applicants and the evaluators are heavily based on the trustworthiness of these 
services.

There exists rich literature comparing these bibliographic databases from the perspectives of source-coverage, citations, author profiles and scientometric 
indicators; the interested reader can check for instance the following 
articles: \cite{Bar-Ilan-Scientometrics08,Archambault-JASIST09,Franceschet-Scientometrics10,Vieira-Scientometrics09,Gusenbauer-Scientometrics19,Martin-JNL-Informetrics18}. 
Especially for Google Scholar, analysis of literature~\cite{Halevi-JNL-Informetrics17} concludes that it currently lacks quality control, clear indexing
guidelines, and it can be easily manipulated~\cite{Lopez-JASIST14}.
However, none of these services' shortcomings recorded by the aforementioned studies have presented any solid evidence that the reported errors can severely
alter the profile of individual articles or even scientists. Thus, these online services continue to enjoy our full confidence.

\begin{figure}[!hbt]
\begin{center}
\includegraphics[scale=.9]{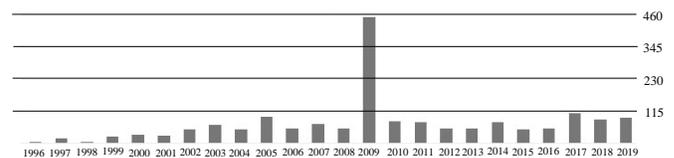}
\end{center}
\caption{A scholar profile implying the existence of supernovae article(s).}
\label{fig-mv-scholar-profile}
\end{figure}

Getting back to this article's question and looking at Figure~\ref{fig-mv-scholar-profile} which is a Google Scholar scientist profile, we could probably be
inclined to answer affirmatively this question; we could argue for instance that one (or a few articles) of that author developed an innovative idea for a problem
which soon got surpassed by another more fit and/or more appropriate idea for that problem. However, this explanation sounds somehow unreasonable knowing that
the articles `live' for a few years (around five) \cite{Galiani-TR17} before their performance (i.e., impact) declines significantly.
It is really bizarre to have an -- almost five times -- {\it abrupt increase in citations within a year which is followed by an immediate return to the previous
low-performance situation}.

So, what is so special with this profile or what is wrong with Google Scholar?
A careful examination of that profile\footnote{\tiny https://scholar.google.com/citations?user=A1mijQ8AAAAJ\&hl=el\&oi=ao} whose peak performance is during 2009,
revealed that the source of this unusual behaviour are four articles; let us call them 
Article1\footnote{\tiny https://scholar.google.com/scholar?oi=bibs\&hl=el\&cites=2799545367351243884},
Article2\footnote{\tiny https://scholar.google.com/scholar?oi=bibs\&hl=el\&cites=14606352744256799080},
Article3\footnote{\tiny https://scholar.google.com/scholar?oi=bibs\&hl=el\&cites=387167649894214057}, and
Article4\footnote{\tiny https://scholar.google.com/scholar?oi=bibs\&hl=el\&cites=6464952970512157755}.
These articles appear to have attracted a very large number of citations from articles appearing in an encyclopedia, namely 
``Encyclopedia of Database Systems''\footnote{https://www.springer.com/gp/book/9780387355443} published by Springer in~2009.
Article1 got~$89$ citations, Article2 got~$87$ citations, Article3 got~$83$ citations, and Article4 got~$83$ citations from this encyclopedia.
These article have the following number of total citations: Article1 has $402$, Article2 has $214$, Article3 has $107$, and Article4 has $105$ citations in total.
Therefore, each of these articles has increased its citation number by an amount $22.14\%$, $40.65\%$, $77.57\%$ and $79.04\%$ respectively due to 
that encyclopedia.

But if so many articles from an encyclopedia cite these particular articles, then why these articles' fame diminished so quickly afterwards?

The question has no scientometric explanation simply because this behaviour is due to a Google Scholar error! This error has dramatic influence on (some of
the) authors of the articles radically changing their profile by faking their true number of citations at a significant percentage.
Thus, the citations from this encyclopedia which are $342(=89+87+83+83)$ in total are fake\footnote{Additionally there exist a couple of legitimate citations to
each of these articles by that encyclopedia, which are actually self-citations.}. How did this happen? This encyclopedia included all entries (i.e., lemmas) 
starting from a specific letter into a single pdf file, so for instance all lemmas starting from the letter `C' are included into the same pdf file. Then all 
lemmas starting from the letter `C' and which appear before the lemma 'Closest Pairs' -- that has true citations to these particular four articles -- are taken 
as citers to them.

However, this `small' error improved this particular author's impact at an amount equal to~$\frac{342}{1787-342}=23.66\%$.
For the first author\footnote{\tiny https://scholar.google.com/citations?user=Q\_Zl0BUAAAAJ\&hl=el\&oi=ao} of these articles, the improvement is even 
bigger, i.e., it is~$\frac{342}{1361-342}=33.5\%$. A similar peak appears in the performance of the third and fourth author for the year 2009, but it is not so
striking because these cites contribute very little to their overall impact. 

Clearly this is not a duplicate citation counting problem. Besides the presence of duplicates in Scholar is rare, only~$0.2-0.3\%$~\cite{Moed-JNL-Informetrics16}.
It is an issue of quality control, i.e., error in linking citations to articles. We can not tell how extensive the problem is because Scholar does not allow for 
unlimited access to its database. A similar problem was reported in~\cite{Moed-JNL-Informetrics16}.

Even though Google Scholar has (by almost no doubt) the widest coverage and highest citation recall~\cite{Gusenbauer-Scientometrics19}, the possible existence 
of many such erroneous records in its database that are dramatically affecting the performance of articles and scientists is a serious issue, which strengthens
the opinion that Google Scholar should be used in conjunction with controlled databases when performing analysis or using bibliometric data for purposes
of funding or promotion.

So, this little hunt for supernovae articles was not successful. However, it succeeded in detecting a serious case of lack of quality control in Google Scholar 
which has dramatic effects on individual articles' and scientists' impact. Unfortunately, we were not able to measure the extent of the problem, since unlimited
computerized access to Scholar's database is not possible.

\section{Conclusions}
\label{sec-conclusions}

The motivation for the work described in this short article was an observation is some Google Scholar's profiles which was never encountered before; although
studies suggest that the articles keep attracting an increasing number of citations per year for the first five years~\cite{Galiani-TR17}, and then their annual
performance drops (usually smoothly), these profiles implied the existence of a kind of {\it supernovae articles}, which presented a large peak in a specific
year and immediately the next year their annual impact dropped (almost) an order of magnitude. However, a closer examination of the data showed that this was 
simply an error in linking citations to articles made by Google Scholar. This error was particularly interesting because, the citing articles were originating
from a premier publisher which has opened its content to bibliometric analysis services. Moreover, this error had dramatic impact altering the overall picture
of both articles and scientists. Therefore, our findings strengthen the opinions of those who call for very careful use of non-controlled bibliographic databases
in research evaluation.

\bibliographystyle{plain}
\bibliography{snovae}

\begin{IEEEbiography}[{\vspace*{-1.1\baselineskip}\includegraphics[height=1in,clip,keepaspectratio]{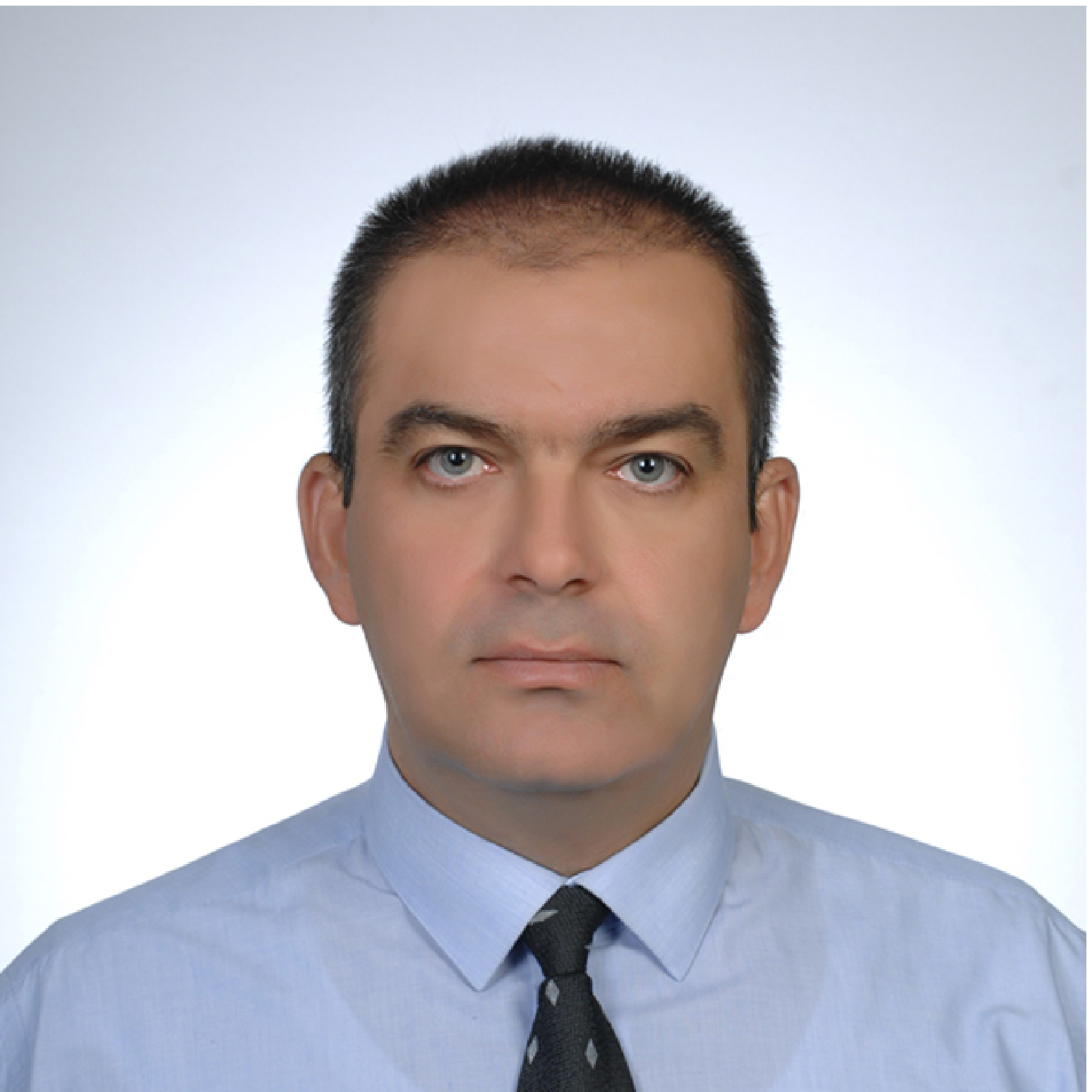}}]{Dimitrios Katsaros}
is associate professor in the Department of Electrical and Computer Engineering at the University of Thessaly. 
More info can be retrieved from https://dana.e-ce.uth.gr/. Katsaros can be reached at dkatsar@e-ce.uth.gr. 
\end{IEEEbiography}

\end{document}